\documentstyle[aps,prl,multicol]{revtex}

\begin{document}
\setlength{\textwidth}{150mm}
\setlength{\textheight}{240mm}
\setlength{\parskip}{2mm}

\input{epsf.tex}
\epsfverbosetrue

\title{Stability criterion for attractive Bose-Einstein condensates}

\author{Luc Berg\'e$^1$, Tristram J. Alexander$^2$, and Yuri S. Kivshar$^2$}

\address{$^1$Commissariat \`a l'Energie Atomique, CEA/Bruy\`eres-le-Ch\^atel, B.P.12, 91680 Bruy\`eres-le-Ch\^atel, France \\
$^2$ Optical Sciences Center, Australian National University, Canberra ACT 0200, Australia}

\maketitle

\normalsize
 
\vspace{1cm}

\begin{abstract}
A general stability criterion is derived for the ground states of the Gross-Pitaevskii equation, which describes attractive Bose-Einstein condensates 
confined in a magnetic trap. These ground states are shown to avoid the collapse
 in finite time and are proven to be stable in two and three spatial dimensions.
\end{abstract}

{\it PACS numbers} : 03.75.Fi, 03.65.Ge, 05.30.Jp, 47.20.Ky, 02.30.Jr
\begin{multicols}{2}
\narrowtext

\section{Introduction}

Experimental observation of Bose-Einstein condensation
(BEC) in ultracold atomic clouds \cite{BEC} has stimulated a new direction
in the study of macroscopic quantum phenomena. Basically, the interaction 
between two confined bosons in a condensate is determined by the s-wave 
scattering length $a_0$ and it can be either repulsive ($a_{0}>0$) or attractive
$(a_{0}<0$). Although first BEC experiments were commonly realized with gases 
promoting a positive scattering length, trapped $^7$Li atom gases, which are 
characterized by a negative scattering length, have raised an increasing 
interest \cite{Li} justified by the rich and complex dynamics mixing 
instability and generation of solitonlike structures, which substantially 
alters the formation of condensates. Furthermore, novel experimental results 
\cite{feshbach} suggested the possibility of using so-called {\em Feshbach 
resonances} to continuously detune $a_0$ from positive to negative values by 
means of an external magnetic field, which brings a new insight into the
experimental realization of BECs with attractive interactions. 

The dynamical behaviors of gases with negative scattering length are especially 
interesting, because in the absence of trapping the condensate is described by 
the nonlinear Schr\"odinger (NLS) equation, whose localized, multi-dimensional 
solutions are unstable and may collapse in finite time (for a review see, e.g., 
Ref.\cite{berge98}). Similarly, collapse also occurs in confined condensates with attractive interactions, as emphasized by many theoretical works 
\cite{pitaevski,watanabe,wadati}, and sequences of collapse events have been 
experimentally detected in BECs of $^7$Li gas \cite{sackett}, in which the 
condensate was observed to shrink on time scales of the trap oscillation period.
General conditions for collapse follow from the virial theorem applied to the 
Gross-Pitaevskii (GP) equation 
\begin{equation}
\label{1}
\displaystyle{i {\hbar} \frac{\partial \psi}{\partial t} = -
\frac{{\hbar}^2}{2m} \nabla^2 \psi + \frac{m \omega^2}{2} (r^2 + \gamma^2
z^2) \psi + U_0 |\psi|^2 \psi,}
\end{equation}
where $\psi$ denotes the BEC wavefunction, $\nabla^2$ is the three-dimensional 
Laplacian, $\gamma$ measures the anisotropy ratio of the condensate and $U_0 
\equiv 4\pi {\hbar}^2 a_{0}/m$ characterizes the two-body interaction ($m$ is 
the atom mass and $\omega$ the trap frequency). These conditions imply the 
existence of a critical value, $N_{\rm cr}$, that the total number of atoms $N 
\equiv \int |\psi|^2 d{\vec r}$ must exceed for initiating the collapse. For 
$N$ below critical, attractive condensates are now known to be capable of 
relaxing to stationary states of the GP equation, reading as $\psi({\vec r},t) 
= \phi({\vec r}) \exp{(-i\mu t/{\hbar})}$, where $\mu$ has the meaning of a 
chemical potential. These solutions consist of solitonlike ground states, whose 
stability was recently analyzed by means of numerical continuation methods 
\cite{huepe}. It was in particular revealed that as $N$ approaches $N_{\rm cr}$ 
from below, the condensate modeled by this ground state becomes more unstable 
and no stationary condensate forms for $N > N_{\rm cr}$. Besides, a new 
investigation \cite{michinel} showed that BEC with large negative scattering 
length can be modulationally unstable and decay into periodic fringe patterns 
of parallel solitons, which again stresses the important role of the latter 
in this field.

In spite of these recent results, some fundamental points, however, remain 
unsolved. First, conditions for the existence of stationary condensates have 
never been established for high-dimensional systems and we still do not know 
how they may be compatible with the conditions for collapse, which mainly 
follow from the Hamiltonian properties of the GP equation. Second, we ignore 
how trapped condensates can evolve towards ground states. Third, for practical
 uses, it is highly desirable to know whether stability of stationary 
condensates can be inferred from their number of particles, {\em and from 
this number only}. The aim of this paper is to answer these questions and to 
clear up the link between the non-occurrence of collapse with respect to the 
stability of localized, stationary ground states when $a_{0} < 0$. In the 
following, we review the conditions required for the collapse of 
two-dimensional (2D) and three-dimensional (3D) condensates and present a 
rigorous derivation of the stability criterion for the ground-state solutions 
of the GP equation. 3D isotropic BECs are analyzed in the limit $\gamma^2 
\rightarrow 1$, for which radially-symmetric solutions only depend on the 
radius $r = \sqrt{x^2 + y^2 + z^2}$ in Eq. (\ref{1}). In 2D, the model (\ref{1})
 can be employed to describe the dynamics of anisotropic condensates having a 
pancake-shape geometry with $\gamma^2 \gg 1$. In this configuration, $\psi = 
\psi'(x,y,t) \phi(z)$ exhibits a longitudinal component, $\phi(z)$, frozen on 
the Gaussian shape $\phi(z) = (\gamma/\pi)^{1/4} \exp (- \gamma z^2/2)$ 
\cite{michinel,Gauck}. The condensate dynamics is then basically governed by 
the 2D version of Eq. (\ref{1}) with radius $r = \sqrt{x^2 + y^2}$. 1D BECs are
 disregarded, since related results are already available in the current literature devoted to optical solitons \cite{sergei}. Instead, we focus our attention on situations allowing for collapse, which basically concerns high 
dimension numbers \cite{LB}. As collapse properties depend on more than the 
dimensionality rather than the geometry of nonlinear objects, we only address 
the stability of isotropic condensates.

\section{Collapse vs. ground states}

>From the viewpoint of the nonlinear dynamics of localized modes, studying 
multi-dimensional trapped condensates is equivalent to investigating the 
solutions of the $D$-dimensional equation
\begin{equation}
\label{5}
\displaystyle{i \partial_t u + \nabla^2 u - \frac{\Omega_0^2}{4} r^2 u + 
|u|^2 u = 0,}
\end{equation}
where $\partial_{t}$ is a partial derivative in time and $\nabla^2 \equiv 
r^{1-D} \partial_r r^{D-1} \partial_r$. In Eq. (\ref{5}), the time and space 
variables are measured in units $(\hbar/m\omega)^{1/2}$ and $2/\omega$, 
respectively; the new wavefunction $u$ scales as $(\hbar \omega/2U_0)^{-1/2}$, 
and $\Omega_0 = 2$ stands for the trap frequency $\omega$, which is formally 
kept undefined for technical convenience. Briefly speaking, collapse occurs as 
the mean-square radius of the wavefunction, $\langle r^2 \rangle \equiv (1/N) 
\int r^2 |u|^2 d{\vec r}$, where $N = \int |u|^2 d{\vec r}$ accounts for the 
rescaled particle number, tends to zero in finite time, while the amplitude of 
$u({\vec r},t)$ diverges in the same limit. To be triggered off, this singular 
phenomenon requires special conditions on the initial data $R_0^2 \equiv \langle
 r^2 \rangle|_{t = 0}$ and $\dot{R}_0^2 \equiv \partial_t \langle r^2 \rangle
|_{t = 0}$, among which we henceforth consider wavefunctions with no initial divergence and set $\dot{R}_0 = 0$, for the sake of clarity. These
requirements are basically inferred from the so-called "virial" relation
\begin{equation}
\label{7}
\displaystyle{\partial_t^2 \langle r^2 \rangle = \frac{8H}{N} + \frac{(4 -
2D)}{N} \int |u|^4 d{\vec r} - 4 \Omega_0^2 \langle r^2 \rangle,}
\end{equation}
which is merely derived by multiplying Eq.(\ref{5}) with $(r^2u^*)$ and 
$({\vec r} \cdot {\nabla u}^*)$ ($^*$ means complex conjugate) and by 
combining the imaginary and real parts of the results, respectively 
\cite{pitaevski,watanabe,wadati,LB}. Equation (\ref{7}) then shows that the 
mean-square radius $\langle r^2(t) \rangle$ inevitably vanishes at a finite 
time, $t_c$, whenever $D \geq 2$ and when the conserved Hamiltonian integral
\begin{equation}
\label{8}
\nonumber
\displaystyle{H \equiv X - Y + \frac{\Omega_0^2}{4} N \langle r^2 \rangle,}
\end{equation}
where
\begin{equation}
\label{8ter}
\displaystyle{X \equiv \int |\nabla u|^2 d{\vec r},\,\,\,\,\,Y \equiv 
\frac{1}{2} \int|u|^4 d{\vec r},}
\end{equation}
fulfils conditions making the right-hand side of (\ref{7}) negative.

Two analytical estimates are usually employed to determine necessary 
requirements on the particle number for initiating the collapse 
\cite{weinstein,Kuz95}, namely, the "uncertainty" inequality
\begin{equation}
\label{8bis}
\displaystyle{N \leq (2/D)^2 X \langle r^2 \rangle,}
\end{equation}
from which the gradient norm $X$ blows up as $\langle r^2 \rangle \to 0$, and 
the Sobolev inequality
\begin{equation}
\label{S}
\displaystyle{Y \leq (2/DN_0)(4/D - 1)^{D/2-1} X^{D/2} N^{2 - D/2}.}
\end{equation}
This expression proceeds from optimizing the estimate $Y \leq C X^{D/2} 
N^{2-D/2}$ by means of variational methods, which amount to minimizing the 
functional $J\{u\} = X^{D/2} N^{2 - D/2}/Y$ and provide a dependence between 
the best constant $C_{\rm best}$ and the quantity $N_0$. Here, $N_0$ is the 
value of $N$ computed with the radially-symmetric ground-state $\chi_0(r)$ 
satisfying the NLS equation $- \chi_0 + \nabla^2 \chi_0 + \chi_0^3 = 0$ with 
no trap, such that $N_0 = 11.68$ for $D = 2$ and $N_0 = 18.94$ for $D = 3$ 
\cite{weinstein}. When $\Omega_0 = 0$ [free NLS], a sufficient condition for 
collapse is $H \leq 0$. In the 2D case, Eq. (\ref{S}) then imposes that $N$ 
must be larger than $N_{\rm cr} = N_0 \simeq 11.7$. In the 3D case, sharper 
estimations of the complete identity (\ref{7}), $\partial_t^2 \langle r^2 
\rangle = (4/N)(3H - X)$, yield the more stringent sufficient condition for 
collapse $H < N_0^2/N$, after bounding $H$ from below with (\ref{S}) 
\cite{Kuz95}.

When $\Omega_0 \neq 0$, the mean-square radius $\langle r^2 \rangle$ 
satisfies \cite{LB}
\begin{equation}
\label{8b}
\nonumber
\displaystyle{\langle r^2 \rangle \leq \left(R_0^2 - \frac{2H}{N 
\Omega_0^2}\right) \cos{(2\Omega_0 t)} + \frac{2H}{N \Omega_0^2},}
\end{equation}
where the strict equality concerns the 2D case only, and $H \leq 0$ is also 
sufficient for collapse, since it leads to $\partial_t^2 \langle r^2 \rangle < 
0$ from Eq. (\ref{7}). Because $H$ expands as $H = H_{\Omega_0 = 0} + H_{\rm 
cr}$, where $H_{\rm cr} \equiv \frac{1}{4} N \Omega_0^2 R_0^2$, this requirement
 implies $H_{\Omega_0 = 0} < 0$. For $D = 2$, $N$ must thus necessarily exceed 
$N_{\rm cr} = N_0$ \cite{pitaevski,LB}. In the opposite case $H > 0$, two 
distinct dynamics can develop: (a) When $H_{\Omega_0 = 0} < 0$, collapse occurs 
again within the domain $H < H_{\rm cr}$ for which the existence of a finite 
maximal blow-up time, $t_c^{\rm max} = (1/2\Omega_0) \cos^{-1}{[1/(1 - 2H_{\rm 
cr}/H)]}$, makes sense. In the special case $N = N_{\rm cr}$ ($H = H_{\rm cr}$),
 2D wavefunctions blow up with an amplitude diverging at $t_c = \pi/2 \Omega_0$. (b) When $H_{\Omega_0 = 0} > 0$ with $N < N_{\rm cr}$, there is no collapse for 
$D = 2$: condensates having $H > H_{\rm cr}$ never blow up and evolve by 
oscillating with the constant frequency $2 \Omega_0$ \cite{gaididei}. For $D = 
3$, the latter conclusion must, however, be subdued, because sharper criteria 
for blow-up may still apply \cite{pitaevski,Kuz95}, so that $H > H_{\rm cr}$ 
just consists of a {\em necessary} condition for the absence of collapse.

Hence, although reinforced with a parabolic trap \cite{LB}, the collapse 
generally manifests under specific conditions, such as $H \leq H_{\rm cr} = 
\frac{1}{4}N\Omega_0^2 R_0^2$. In the opposite case, $H > H_{\rm cr}$, collapse 
can disappear and the wavefunction $u$ does not
systematically spread out, but may instead oscillate or even form {\em robust} 
solitonlike states. {\em This possibility of forming stable condensates remains 
an open problem for high dimension numbers}. Therefore, it is worth knowing 
whether the stationary bound states of Eq. (\ref{5}), expressing as
\begin{equation}
\label{BS}
\displaystyle{u({\vec r},t) = \chi({\vec r}) \exp{(i \Lambda t)},}
\end{equation}
can exist for spatial dimensions $D \geq 2$. These ground states are defined by 
the solutions of the differential equation
\begin{equation}
\label{10}
\displaystyle{- \Lambda \chi + \nabla^2 \chi -
\frac{\Omega_0^2}{4} r^2 \chi + |\chi|^2 \chi  = 0,}
\end{equation}
which functionally depend on $\Lambda$ and are localized in space, i.e., 
$\chi({\vec r}) = 0$ at $r \rightarrow +\infty$. Here, the parameter $\Lambda$ 
is sign-opposite to the energy eigenvalue $E = - \Lambda$, which corresponds to 
the above-defined chemical potential $\mu$ in reduced units. To start with, we examine the precise conditions under which
these states may form. We multiply Eq. (\ref{10}) by $\chi^*$ and ${\vec r} 
\cdot {\vec \nabla} \chi^*$, then combine the results to find
\begin{equation}
\label{13}
\nonumber
\begin{array}{l}
\displaystyle{X_s = \frac{N_s}{(4-D)}
[D \Lambda + \frac{\Omega_0^2}{4} (D + 4) \langle r^2 \rangle_s],} \\
*[9pt]
{\displaystyle Y_s = \frac{N_s}{(4-D)} [2\Lambda + \Omega_0^2 \langle r^2 
\rangle_s],}
\end{array}
\end{equation}
where subscript $s$ applies to the integrals computed on the ground-state 
solution. As $X_s$ and $Y_s$ are both positive quantities, it is easy to check 
that localized solutions $\chi$ exist for $D < 4$, provided that
\begin{equation}
\label{14}
\displaystyle{\Lambda > - \frac{\Omega_0^2}{2} \langle r^2 \rangle_s.}
\end{equation}
Unlike their free NLS counterparts ($\Omega_0 = 0$), {\em confined stationary 
states can thus exist for negative parameters} $\Lambda$ and the requirement 
(\ref{14}) implies that the energy $E$ must be less than twice the mean-square 
radius of the stationary condensates. Furthermore, the integral $H$ for the ground states reads
\begin{equation}
\label{15}
\nonumber
\displaystyle{H_s = \frac{D-2}{4-D} \Lambda N_s + \frac{\Omega_0^2}{4-D}
N_s \langle r^2 \rangle_s,}
\end{equation}
which ensures $\partial_t^2 \langle r^2 \rangle_s = 0$ from Eq. (\ref{7}), as 
expected. Combining relation (\ref{15}) with the existence condition (\ref{14}) 
then yields the constraint that the Hamiltonian must satisfy on the stationary 
solutions:
\begin{equation}
\label{16}
\displaystyle{H_s \geq \frac{\Omega_0^2}{2} N_s \langle r^2 \rangle_s > 0.}
\end{equation}
Thus, $H_s$ for steady solutions of trapped BECs is strictly positive. Moreover,
 we can observe from the result (\ref{16}) that $H_s$ {\em belongs to the range}
 $H_s > \frac{1}{4}N \Omega_0^2 \langle r^2 \rangle_s$, {\em where
there is no collapse for $D = 2$ if $N_s < N_{\rm cr}$, and where collapse can 
be prevented for $D = 3$}. On the other hand, since collapse implies the 
blow-up of the gradient norm $X \rightarrow +\infty$ as $\langle r^2 \rangle 
\rightarrow 0$ [see Eq.(\ref{8bis})], a sharper condition assuring the collapse 
at dimensions $D > 2$ can be inferred in the domain $X > X_s$ from the virial 
relation (\ref{7}) rewritten as
$$
\partial_t^2 \langle r^2 \rangle = \frac{4D}{N}[H - (1-2/D)X] - (D+2) \Omega_0^2 \langle r^2 \rangle,
$$
and it reads $H \leq (1 - 2/D) X_s$. From expression (\ref{15}), $H_s$ can be 
seen to never belong to this class of Hamiltonians. Therefore, as $H_s$ always 
lies above this bound together with the limit value $H_{\rm cr} \equiv 
\frac{1}{4}N\Omega_0^2 R_0^2$ below which any initial datum certainly promotes 
the collapse, it is important to investigate the stability of the stationary 
states $\chi$.

\section{Stability criterion}

We first use the heuristic argument, following which the functional
dependences of $H$ on $u$ under the constraint of a fixed particle number
$N$ provides some qualitative information about stability. Indeed, as is
readily seen from the variational problem $\delta(H + \Lambda N) = 0$, $\chi$ 
governed by (\ref{10}) realizes an extremum
of $H$ at fixed $N$. Under this constraint, we employ the compatible
substitution $u = a^{-D/2} \chi(r/a)$, where $a$ plays the role of a
Lagrange multiplier, and we plug it into $H$ to obtain
\begin{equation}
\label{17}
\nonumber
\displaystyle{H_a = X_s/a^2 - Y_s/a^D + a^2 I_3,}
\end{equation}
where the integrals $X_s$ and $Y_s$ are here expressed in terms of $\xi = r/a$, 
while $I_3 \equiv \frac{1}{4}\Omega_0^2 \int \xi^2 |\chi(\xi)|^2
d{\vec \xi}$. It is then clear from Eq. (\ref{17}) that $H_a$ admits a
global minimum for $D = 2$, whenever $X_s > Y_s$. Also for $D = 3$, $H_a$
may exhibit a local minimum under the same condition. The minima of $H_a$
are given by the roots of the identity $\delta H_a/\delta a|_{a = 1} = 0$,
i.e., $X_s - DY_s/2 - I_3 = 0$, which is nothing else but the
characteristic relation for the ground states, as this can be refound from Eq. 
(\ref{13}). This argument points out
that the minima of $H$ are reached on the stationary states $\chi$, that
constitute stable equilibrium solutions around which any nearby
solution can be trapped, provided that $X_s > Y_s$. Note that this
inequality is always satisfied by all bound states $\chi$ when $D \geq 2$. For 
comparison, with the same arguments the free NLS ground states
($\Omega_0 = 0$) are found to be unstable and $H_a$ has no global minimum
for these dimension numbers.

To derive a more rigorous proof for the stability of stationary
wavefunctions, we now perturb the latter by means of the following solution
\begin{equation}
\label{20}
\displaystyle{u(r,t) = [\chi(r) + v(r,t) + i w(r,t)] \mbox{e}^{i\Lambda t},}
\end{equation}
where $v,w \ll \chi$ are real functions, which we assume to be
radially-symmetric. Inserting (\ref{20}) into Eq. (\ref{5}), linearizing and 
decomposing the resulting equation into real and
imaginary parts then yield
\begin{equation}
\label{SP}
\displaystyle{L_0 w = \partial_t v, \,\,\,L_1 v = - \partial_t w,}
\end{equation}
where the differential operators
\begin{equation}
\label{VK}
\displaystyle{L_0 = - \nabla^2 + \Lambda - \chi^2 + \frac{\Omega_0^2}{4} r^2, 
\,\,\,L_1 = L_0 - 2 \chi^2,}
\end{equation}
have the properties
\begin{equation}
\label{VK2}
\displaystyle{L_0 \chi = 0, \,\,\,\,\,L_1 \partial \chi/\partial \Lambda = - 
\chi,}
\end{equation}
\begin{equation}
\label{VK3}
\displaystyle{L_1 {\vec \nabla} \chi = - \Omega_0^2 {\vec r} 
\chi/2,\,\,\,\,\,L_0 {\vec r} \chi = - 2 {\vec \nabla} \chi.}
\end{equation}
with ${\vec \nabla} \chi \equiv ({\vec r}/r) \partial_r \chi$. It is interesting
 to notice that the small-amplitude perturbations described by Eqs. (\ref{20}) to (\ref{VK}) provide an alternate form of the usual Bogoliubov excitations. 
Here, $\chi$ is supposed to be the unique, radially-symmetric solitonlike
bound state of Eq. (\ref{10}) with no zeroes. So, as $L_0 \equiv -
\chi^{-1} \nabla [\chi^2 \nabla (1/\chi)]$ is nonnegative and admits the
discrete eigenvalue $0$ for $\chi$ only, zero is the lowest
eigenvalue of $L_0$. In addition, since $\chi$ has no node, $\nabla \chi$
possesses a single node. We then apply a basic theorem from spectral theory 
claiming that if $\psi_k$ is an eigenstate for $L_1 < L_0$ such that $L_1 
\psi_k = \lambda_k \psi_k$, and if $\psi_k$ has exactly $k$ zeroes, then 
$\lambda_k$ is the $(k+1)$th eigenvalue ranked as $\lambda_0 < \lambda_1 < ... 
<\lambda_k$. This theorem holds when the eigenfunctions of $L_1$ are 
$L^2$-integrable, which is here ensured since the linear eigenstates of $L_1$ 
decay much faster with Gaussian tails than their free NLS counterparts. For 
$\Omega_0 = 0$, $\lambda_1$ is equal to zero and, therefore, a unique negative 
eigenvalue $\lambda_0 < 0$ exists. Let us now observe what happens with a 
trapping potential ($\Omega_0 \neq 0$). First, $L_1$ has certainly a
strictly positive eigenvalue since
\begin{equation}
\label{22}
\nonumber
\displaystyle{<\nabla \chi | L_1 \nabla \chi > = \frac{D \Omega_0^2}{4} N_s
> 0,}
\end{equation}
where $<|>$ denotes the standard $L^2$ scalar product applied to real
functions. Second, it is obvious that the vectors ${\vec \nabla} \chi$ and $- 
{\vec r} \chi$ are both orthogonal to the ground state $\chi$. A combination of Eq. (\ref{VK3}) then leads to
\begin{equation}
\label{22bis}
\displaystyle{L_0 L_1 {\vec \nabla} \chi = \Omega_0^2 {\vec \nabla} \chi}
\end{equation}
and, thus, ${\vec \nabla} \chi$ is an eigenstate of $L_0L_1 < L_0^2$ with 
positive eigenvalue. For $L_0$ being positive definite, this property indicates 
that $\lambda_1 > 0$. The lowest eigenvalue $\lambda_0$ of $L_1$ is necessarily 
negative, although larger than for $\Omega_0 = 0$, because $<\eta|L_1 \eta>$ 
attains negative values
with, e.g., $\eta = \chi$. Moreover, $L_1$ has exactly one negative eigenvalue 
in its spectrum \cite{rose}. These technical conditions being fulfilled, we 
look for perturbations growing as $v,w \sim \mbox{e}^{\Gamma t}$ with growth 
rate $\Gamma$ and determine the condition assuring the stability of ground 
states by maximizing
\begin{equation}
\label{22b}
\nonumber
\displaystyle{\Gamma^2 = - \frac{<v|L_1 v>}{<v|L_0^{-1} v>}}
\end{equation}
in the subspace of functions $v$ orthogonal to $\chi$. Stability then
results from the proof that $<v|L_1 v>$ is positive under the constraint
$<v|\chi> = 0$. According to standard procedures \cite{vakhitov,kolokolov,VK}, 
we identify the discrete spectrum of $L_1$ by setting $L_1 v = {\bar \lambda} v 
+ \alpha \chi$, where $\alpha$ is a Lagrange multiplier related to the
orthogonality condition $<v|\chi> = 0$. We construct a complete basis of
orthonormalized eigenfunctions as $|v> = \sum_n C_n |\psi_n>$, such that
$L_1 |\psi_n> = \lambda_n |\psi_n>$, and combine these relations to get
\begin{equation}
\label{22c}
\nonumber
\displaystyle{G({\bar \lambda}) = \sum_n \frac{<\chi|\psi_n><\psi_n|\chi>}{\lambda_n -
{\bar \lambda}} = 0.}
\end{equation}
In (\ref{22c}), the function $G({\bar \lambda})$ monotonously increases from 
$-\infty$ to $+\infty$ in the range $]\lambda_0,\lambda_1[$. With
$\lambda_0 < 0$ and $\lambda_1 > 0$, stability follows from the sign of
$G(0) \equiv <\chi | L_1^{-1} \chi>$, which must be negative in order
to ensure ${\bar \lambda} > 0$. As $G(0)$ is defined by $G(0) \equiv -
<\chi | \partial \chi/\partial \Lambda>$, we conclude that {\em a
sufficient condition for stability} is given by
\begin{equation}
\label{25}
\displaystyle{\frac{d N_s}{d \Lambda} > 0.}
\end{equation}
This condition guarantees the orbital stability of the stationary condensate. 
By "orbital" stability, it is meant that, modulo the elementary symmetries of 
Eq. (\ref{5}), the shape of a BEC soliton is preserved by perturbations having 
no growing modes, when they act on any solution staying nearby the ground state
 orbit \cite{rose}. Conversely, $dN_s/d\Lambda < 0$ is sufficient for any bound 
state of Eq.(\ref{5}) to be unstable. If we moreover admit that $dN_s/d\Lambda 
= 0$ also leads to instability, condition (\ref{25}) is not only sufficient, 
but also necessary for the stability of ground states.
\section{Numerical results}
\vspace*{-3mm}
Identified through numerical integrations of Eq. (\ref{10}), some 
radially-symmetric, bell-shaped solutions $\chi$ are presented in Figure 1 for 
different parameters $\Lambda$ and $\Omega_0 = 2$. Two-dimensional and 
three-dimensional ground states are illustrated in Fig. 1(a) and Fig. 1(b), respectively.
\begin{figure}
\setlength{\epsfxsize}{9cm}
\epsffile{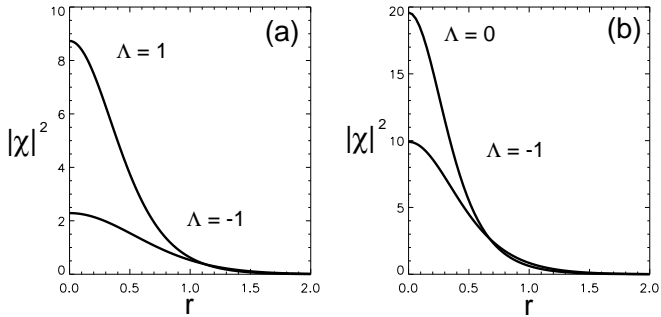}
\vspace{2mm}
\caption{Examples of radially-symmetric, localized stationary
  condensates with different $\Lambda$ and $\Omega_0 = 2$, at both
  dimensions $D=2$ (a) and $D=3$ (b).}
\end{figure}
The dependence $N_s$ versus $\Lambda$ is shown in Figure 2 for condensates 
defined at dimension numbers two [Fig. 2(a)] and three [Fig. 2(b)], with 
$\Omega_0 = 2$. Dotted lines indicate the variations of the ground-state 
particle number $N_s$ with respect to $\Lambda$ for the free NLS equation 
($\Omega_0 = 0$), namely, $N_s = 11.68$ in 2D and $N_s = 18.94/\sqrt{\Lambda}$ 
in 3D \cite{Kuz95}. Note that $N_s$ for $\Omega_0 \neq 0$ always lies below 
these free particle numbers. In the limit $\Lambda \to +\infty$, the solution 
$\chi$ behaves as free NLS solitons, {\em which are all unstable with 
$dN_s/d\Lambda \leq 0$}. In contrast, as can be seen from this set of figures, 
{\em trapped ground states are stable} for every $\Lambda$ with
moderate values when $D = 2$, whereas for $D = 3$, the criterion (\ref{25}) predicts their 
stability for negative $\Lambda$ only, i.e., $\Lambda < \Lambda_{\rm cr}$, 
where $\Lambda_{\rm cr} \simeq -0.72$ corresponds to the maximum number 
$N_s^{\rm max} \simeq 14.45$.
\begin{figure}
\setlength{\epsfxsize}{9cm}
\epsffile{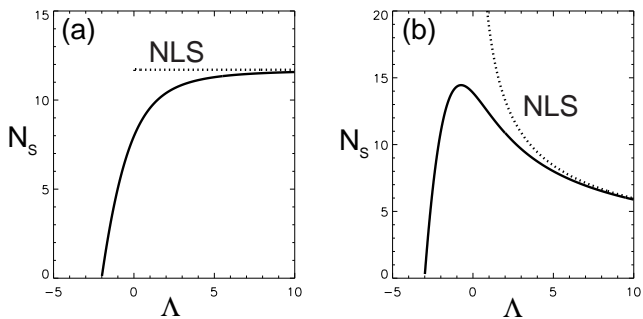}
\vspace{2mm}
\caption{$N_s$ vs. $\Lambda$ for the ground states of Eq. (2).  Solid
  curves refer to 2D (a) and 3D (b) confined ground states.  Dotted
  lines indicate their free NLS limits.}
\end{figure}
The difference between "free" and trapped solutions is that {\em the discrete 
spectrum of $L_1 = L_1^{\Omega_0 = 0} + \Omega_0^2r^2/4$ is here shifted towards
 the range of positive eigenvalues}. Thereby, stability of ground states seems 
"reinforced" by the parabolic trap, compared with the case $\Omega_0 = 0$. Such 
a result cannot be obtained for a detrapping potential ($\Omega_0^2 \rightarrow 
- \Omega_0^2$), because the linear eigenstates of $L_1$ are not $L^2$-integrable
 in this configuration. Even if the above procedure could be applied in that 
case, Eq.(\ref{22}) furthermore suggests that at least two negative eigenvalues 
would appear in the discrete spectrum of $L_1$, from which instability surely 
follows.

Finally, we have plotted in Fig. 3 the temporal evolutions of solutions to Eq. 
(\ref{5}) with a parabolic trap for various initial data. In the 2D case, Fig. 
3(a) shows $|u(0,t)|^2$ for initial conditions defined by the ground states 
$\chi$ with $\Lambda = 1$ and $\Lambda = 5$ for $\Omega_0 = 2$, and with 
$\Lambda = 1$ for $\Omega_0 = 0$. In the 3D case, Fig. 3(b) represents similar 
evolutions from ground states having $\Lambda = -2,-1,0$, among which the 
solution initiated with $\Lambda = 0$ inexorably collapses. Dots show the same 
initial data undergoing small-amplitude periodic perturbations. From those, we 
can observe the robustness of the 2D and 3D ground states of the GP equation, 
for which both inequalities (\ref{16}) and (\ref{25}) are always verified. Note 
that perturbations do not affect the ground state orbits, as long as $\Lambda$ 
is far below the critical values beyond which stationary solutions must become 
unstable, i.e., $\Lambda \gg 5$ for $D = 2$ and $\Lambda > - 0.72$ for $D = 3$. 
Stability is lost as $\Lambda$ attains large positive values, in accordance 
with the criterion (\ref{25}).
\begin{figure}
\setlength{\epsfxsize}{9cm}
\epsffile{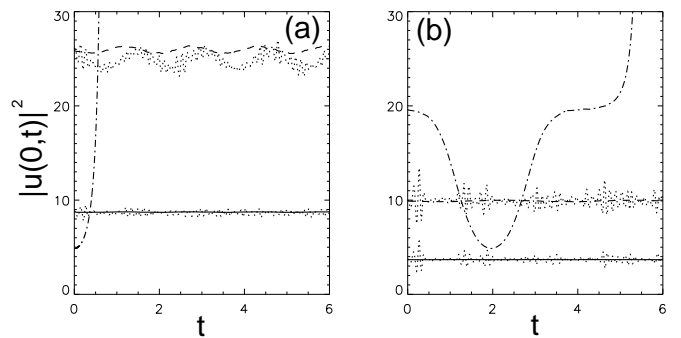}
\vspace{2mm}
\caption{Temporal evolution of $|u(0,t)|^2$ for different initial data with $N = N_s$. (a) $D=2,\,\Lambda = 1$ (solid curve), $\Lambda = 5$ (dashed) for $\Omega_0=2$, and $\Lambda = 1$ for $\Omega_0 = 0$ (dash-dotted). This datum promotes a finite-time collapse at $t_c = \pi/2\Omega_0$. (b) $D=3,\,\Lambda = -2$ (solid), $\Lambda = -1$ (dashed), and $\Lambda = 0$ (dash-dotted) for $\Omega_0=2$. The dots illustrate the action of perturbations introduced at $t = 0$ around the ground states.}
\end{figure}
\section{Conclusion}
We have discriminated the different regions of collapse/no-collapse for the 
non-stationary solutions of Eq. (\ref{5}) and determined the exact conditions 
under which stationary localized states of this equation can exist. We have also demonstrated that the stationary ground states lie in the region $H > H_{\rm cr} > 0$, where collapse is definitively absent in 2D and can be avoided in 3D, and that they are stable provided $dN_s/d\Lambda$ is positive. This result displays 
evidence that a parabolic trap makes the multi-dimensional, stationary solitary 
modes of the GP equation stable for a wide class of parameters accessible in BEC physics.

To conclude, the criterion (\ref{25}) allows for determining straightforwardly 
the stability of condensates from their particle numbers only, which should be 
useful in current experiments.

\end{multicols}


\begin{references}

\bibitem{BEC} M.H. Anderson, J.R. Ensher, M.R. Mattews, C.E. Wieman, and E.A. 
Cornell, {\em Science} {\bf 269}, 198 (1995); K.B. Davis, M.O. Mewes, M.R. 
Andrews, N.J. Van Druten, D.S. Durfee, D.M. Kurn, and W. Ketterle, {\em Phys. 
Rev. Lett.} {\bf 75}, 3969 (1995); see also for review: F. Dalfovo, S. Giorgini, L.P. Pitaevskii, and S. Stringari, {\em Rev. Mod. Phys.} {\bf 71}, 463 (1999).

\bibitem{Li} C.C. Bradley, C.A. Sackett, J.J. Tollet, and R.G. Hulet, {\em Phys. Rev. Lett.} {\bf 75}, 1687 (1995); C.C. Bradley, C.A. Sackett, and R.G. Hulet, 
{\bf 78}, 985 (1997); C.A. Sackett, C.C. Bradley, M. Welling, and R.G. Hulet,
 {\em Appl. Phys. B} {\bf 65}, 433 (1997); C.A. Sackett, H.T.C. Stoof, and R.G. 
Hulet, {\em Phys. Rev. Lett.} {\bf 80}, 2031 (1998); Yu. Kagan, A.E. Muryshev, 
and G.V. Shlyapnikov, {\em Phys. Rev. Lett.} {\bf 81}, 933 (1998).

\bibitem{feshbach} S. Inouye, M.R. Andrews, J. Stenger, H.J. Miesner, D.M. 
Stamper-Kurn, and W. Ketterle, {\em Nature} {\bf 392}, 151 (1998).

\bibitem{berge98} L. Berg\'e, {\em Phys. Rep.} {\bf 303}, 259 (1998).

\bibitem{pitaevski} L.P. Pitaevskii, {\em Phys. Lett. A} {\bf 221}, 14 (1996).

\bibitem{watanabe} K. Watanabe, T. Mukai and T. Mukai, {\em Phys. Rev. A} {\bf 
55}, 3639 (1997).

\bibitem{wadati} T. Tsurumi and M. Wadati, {\em J. Phys. Soc. Jpn.} {\bf 66}, 
3035 (1997); {\em ibid.} {\bf 68}, 1531 (1999); M. Wadati and T. Tsurumi, {\em 
Phys. Lett. A} {\bf 247}, 287 (1998).

\bibitem{sackett} C.A. Sackett, J.M. Gerton, M. Welling, and R.G. Hulet, {\em 
Phys. Rev. Lett.} {\bf 82}, 876 (1999).

\bibitem{huepe} C. Huepe, S. M\'etens, G. Dewel, P. Borckmans, and M.E. Brachet, {\em Phys. Rev. Lett.} {\bf 82}, 1616 (1999).

\bibitem{michinel} H. Michinel, V.M. P\'erez-Garc\'ia, and R. de la Fuente {\em 
Phys. Rev. A} {\bf 60}, 1513 (1999).

\bibitem{Gauck} H. Gauck, M. Hartl, D. Schneble, H. Schnitzler, T. Pfau, and J. 
Mlynek, {\em Phys. Rev. Lett.} {\bf 81}, 5298 (1998); A.I. Safonov, S.A. 
Vasilyev, I.S. Yasnikov, I.I. Lukashevich, and S. Jaakkola, {\em Phys. Rev. 
Lett.} {\bf 81}, 4545 (1998).

\bibitem{sergei} S.K. Turitsyn, {\em Phys. Rev. E} {\bf 56}, R3784 (1997).

\bibitem{LB} L. Berg\'e, {\em Phys. Plasmas} {\bf 4}, 1227 (1997).

\bibitem{weinstein} M.I. Weinstein, {\em Commun. Math. Phys.} {\bf 87}, 567 
(1983).

\bibitem{Kuz95} E.A. Kuznetsov, J. Juul Rasmussen, K. Rypdal, and S.K. Turitsyn,
 {\em Physica D} {\bf 87}, 273 (1995).

\bibitem{gaididei} Yu.B. Gaididei, K.{\O}. Rasmussen, and P.L. Christiansen, 
{\em Phys. Rev. E} {\bf 52}, 2951 (1995).

\bibitem{rose} H.A. Rose and M.I. Weinstein, {\em Physica D} {\bf 30}, 207 
(1988).

\bibitem{vakhitov} N.G. Vakhitov and A.A. Kolokolov, {\em Izv. Vuz. Radiofiz.} 
{\bf 16}, 1020 (1973) [{\em Radiophys. and Quantum Electronics} {\bf 16}, 783 
(1975)].

\bibitem{kolokolov} A.A. Kolokolov, {\em Izv. Vuz. Radiofiz.} {\bf 17}, 1332 
(1974) [{\em Radiophys. and Quantum Electronics} {\bf 17}, 1016 (1976)].

\bibitem{VK} E.A. Kuznetsov, A.M. Rubenchik and V.E. Zakharov, {\em Phys. Rep.} 
{\bf 142}, 103 (1986). 

\end{references}
\end{document}